# Giant saturation magnetization effect in epitaxial $Fe_{16}N_2$ thin films grown on MgO (001) substrate

Nian Ji[1,2,*], Valeria Lauter[3,*], Hailemariam Ambaye[3] and Jian-Ping Wang[1,2,†]

[1]*The Center for Micromagnetics and Information Technologies (MINT) and Department of Electrical and Computer Engineering, University of Minnesota, 200 Union St SE, Minneapolis, Minnesota 55455, USA*

[2]*School of Physics and Astronomy, University of Minnesota, Minneapolis, Minnesota 55455, USA*

[3]*Neutron Scattering Science Division, Oak Ridge National Laboratory, Oak Ridge, TN 37831 USA*

**Abstract**

Whether α''-$Fe_{16}N_2$ possesses a giant saturation magnetization ($M_s$) has been a daunting problem among magnetic researchers for almost 40 years, mainly due to the unshakable faith of famous Slater-Pauling (SP) curve and poor consistency on evaluating its $M_s$. Here we demonstrate that, using epitaxy and mis-fit strain imposed by an underlying substrate, the in-plane lattice constant of $Fe_{16}N_2$ thin films can be fine tuned to create favorable conditions for exceptionally large saturation magnetization. Combined study using polarized neutron reflectometry and X-ray diffraction shows that with increasing strain at the interface the $M_s$ of these film can be changed over a broad range, from ~2.1T (non-high $M_s$) up to ~3.1T (high $M_s$). We suggest that the equilibrium in-plane lattice constant of $Fe_{16}N_2$ sits in the vicinity of the spin crossover point, in which a transition between low spin to high spin configuration of Fe sites can be realized with sensitive adjustment of crystal structure.

In 1972, Kim and Takahashi[1] discovered that iron nitride thin films prepared by evaporating Fe onto glass substrate in an $N_2$ atmosphere possess a giant saturation magnetization ($M_s$) of 2.64T (at RT), which substantially exceeds the known limit ($Fe_{65}Co_{35}$ with $M_s \sim 2.45T$) as predicted by the famous Slater-Pauling (SP) Curve[2, 3, 4]. They attributed the formation of this giant $M_s$ to be due to the presence of α"-$Fe_{16}N_2$, from which they deducted an $M_s$ of 2.76T for that phase, corresponding to an average magnetic moment of 3.0 $\mu_B$/Fe. However, this report went mostly unnoticed due to the unsuccessful synthesis of bulk sample with pure phase given the metastable nature of this phase[5, 6] and the well known difficulties on precisely measuring the $M_s$ of thin films that containing multiple phases. In 1990s', Sugita and co-workers reported a remarkable breakthrough[7, 8, 9, 10] that by introducing epitaxial constrain, pure phase single crystal α"-$Fe_{16}N_2$ can be fabricated using molecular beam epitaxy (MBE) approach and the $M_s$ measured on these samples repeatedly reaches 2.8~3.1T. Following this claim, enormous efforts were dedicated to reproduce their results[11]. However, the reported $M_s$ values on samples produced by different groups cover a disappointingly broad range though epitaxial growth was employed, which is known to stabilize the material metastablilty. The magnetization obtained from these samples still varies from high-$M_s$ of 2.9T[12], intermediate-$M_s$ of 2.6T[13], to non-high $M_s$ of 2.1~2.3T[14, 15, 16]. Until now, the question on whether this material possesses a giant magnetization remains to be a mystery.

One of a problem for epitaxial thin films is a delicate choice of substrate or underlayers. From one side their magnetic contributions are subtle to assess based on conventional magnetometer (VSM or SQUID) methods, which only allow for the evaluation of the average $M_s$ of the entire sample. From the other side the epitaxial growth with a certain mis-match can be used to fabricate thin film systems with new physical properties. In the present work we show that the mis-fit stain introduced with the epitaxial growth induces giant saturation magnetization in of $Fe_{16}N_2$ thin films grown on MgO substrates. We performed Polarized Neutron Reflectometry (PNR) experiments, which for the first time provide the *direct* evidence for the existence of the giant saturation magnetization in $Fe_{16}N_2$ on MgO substrates. The combination of x-ray and polarized neutron reflectometry allows the unambiguous determination of the depth dependent magnetic structure within sub nm resolution[17], and serves as an ideal tool to probe the absolute magnetization of the iron nitride thin film system.

The essential idea of controlling the stain is schematically shown in Fig. 1 that we grow $Fe_{16}N_2$ epitaxially on Fe buffered MgO substrate. It is known that the lattice mis-match between $Fe_{16}N_2$ and Fe is about ~0.3%, which is much smaller than that of Fe and MgO (4.2%). Therefore, by tuning the thickness of Fe buffer, it is possible to the fine-tuning of the in-plane lattice as well as the strain in the $Fe_{16}N_2$. This epitaxial structure is fabricated by facing target sputtering method[18]. Consistent with our previous results on GaAs based samples[19], to provide (001) epitaxy, Fe underlayer with (001) orientation is deposited at 300 º C, which were proven to facilitate the (001) texture. The Fe-N layer is subsequently synthesized by vaporizing iron target using

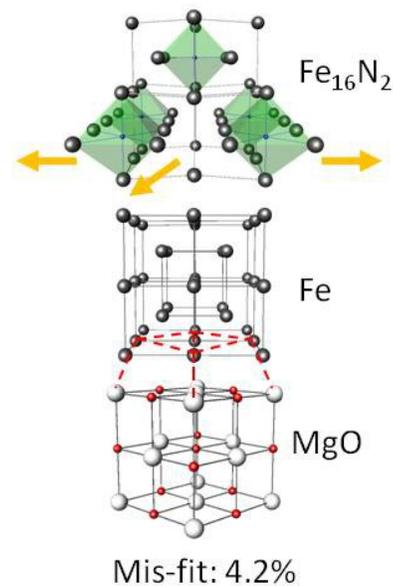

FIG. 1 **A sketch of epitaxial relationship of $Fe_{16}N_2$/Fe/MgO with (001) orientation.** Due to the lattice mis-fit, the $Fe_{16}N_2$ is under tensile strain, which can be tuned by adjusting the thickness of Fe buffer.

thoroughly mixed Ar+$N_2$ plasma with a $N_2$ partial pressure of ~0.35mTorr. The as-deposited samples show a body center tetragonal N-disordered Fe-N martensite with (001) orientation and a stoichiometry close to Fe/N~8/1 according to X-ray diffraction (XRD) and Auger electron spectroscopy analysis. Post-annealing the as-deposited Fe-N samples 40 hrs gives rise to the formation of chemically ordered α"-$Fe_{16}N_2$. Throughout the paper, all the structural and magnetic characterizations are performed on two samples labeled as S1 and S2, with nominal structures of

S1: *Fe-N (40nm)/ Fe (2nm)/ MgO*

S2: *Fe-N (40nm)/ Fe (20nm)/ MgO*

The θ-2θ XRD scan performed on a D5005 diffractometer with Cu Kα source was shown in Fig. 2a. It is seen that both samples exhibit similar (00*l*) orientation, with an out-of-plane lattice constant of 6.28Å. Other than diffractions from substrate and buffer layer Fe (002), the peaks developed at 58.8° and 28.6° can be indexed to $Fe_{16}N_2$ (004) and $Fe_{16}N_2$ (002), respectively. Consistent with previous reports on similar samples (Ref. 13 and 14) that the integrated area ratio of diffraction peaks from (002) and (004) planes does not reproduce what would be expect from single crystal sample, corresponding to a degree of N site ordering (D, as defined in Ref.) in range of 0.26±0.15 for both samples. This is likely due to the relatively short range ordering of the (002) diffraction in contrast to that of (004) diffraction, which can be estimated from the rocking curve measurement (Fig. 2b) that the mosaic spreading of the (002) peak (Δθ~1.6°) is three times as large as that of (004) peak (Δθ ~0.5°).

To verify the idea of strain control as discussed in Fig. 1. In-plane x-ray diffraction with grazing incident geometry was performed on both samples and plotted in Fig. 2c. Aligning the scattering vector along the MgO (*200*), the observed peak in the neighborhood of 44.8 º corresponds to Fe$_{16}$N$_2$ (220) and is coherent with the underlayer Fe (110). It is clear that the peak (Fe$_{16}$N$_2$(220)/Fe(110)) from thicker sample (S2) shows notable shift toward the higher angle as oppose to the thinner one (S1), suggesting the longer average in-plane lattice constant in S1 than that in S2. Typical Φ scan (Fig. 2d) shows regularly distanced four peaks with 90 º spacing, suggesting an expected 4-fold cubic symmetry with the full width of half maximum (FWHM) of 1.16° (Inset in Fig.

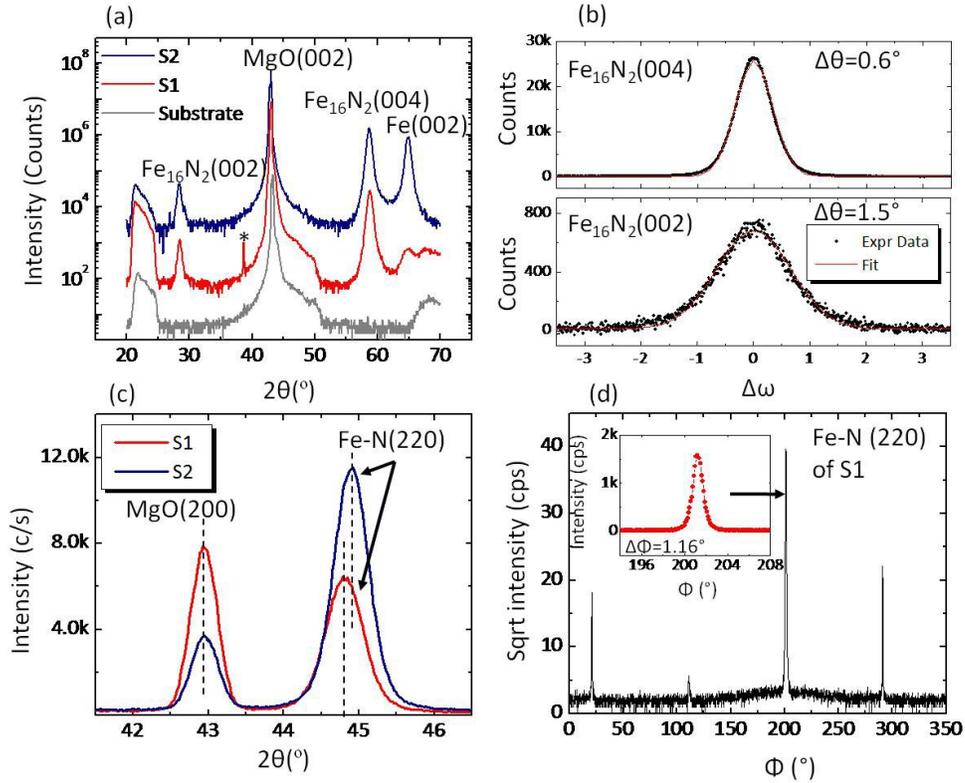

FIG. 2 **Structural characterization using X-ray diffraction.** (a) High angle x-ray diffraction data on sample S1: Fe-N (40nm)/Fe (2nm)/MgO and S2: Fe-N (40nm)/Fe (20nm)/MgO. The peak labeled by (*) comes from the CuK$\alpha_2$. (b) Gaussian fitted Rocking curves measured on Fe$_{16}$N$_2$ (002) and (004) of sample S1 respectively. (c) Grazing incident x-ray diffraction with scattering vector aligned along MgO (2 0 0) on sample S1 and S2. The shift of the peak position upon underlayer thickness increase suggests the tensile strain developed at the bottom interface of the films. Dashed lines are guides to the eye. (d) Φ scan of Fe16N2 with scattering vector aligned on 2θ=Fe$_{16}$N$_2$ (220) of sample S1. The inset shows the zoom-in look of one peak (outlined by arrow).

2d), which is comparable to monocrystalline samples. The combination of out-of-plane and in-plane x-ray diffraction analysis reveals that the prepared films possess body center tetragonal crystal structure (a=5.72Å, c=6.28Å), which coherently follow the epitaxy of MgO substrate with a substantial tensile strain developed at the bottom interface between the film and the substrate.

Low angle x-ray reflectivity curves are shown in Fig. 3a, which are collected from a

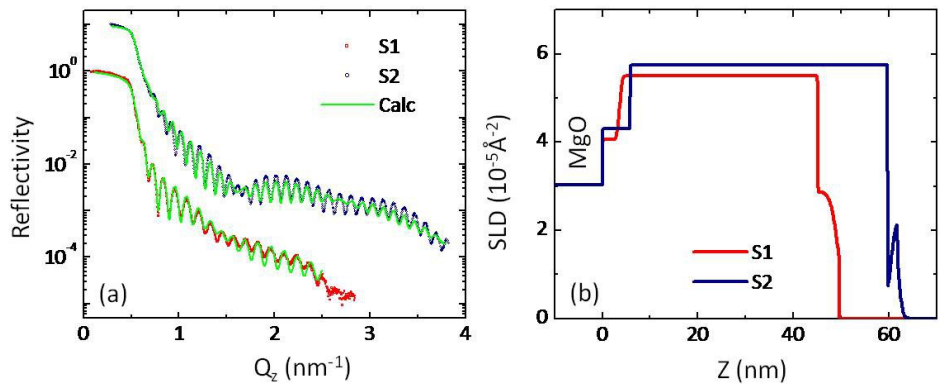

FIG. 3 **X-ray reflectivity characterization**. (a)The fitted x-ray reflectivity curves measured on sample S1 and S2 (vertically offset by a factor of 10. (b) Calculated depth-dependent x-ray scattering length density profiles.

Phillip Pro X'pert x-ray diffractometer with Cu Kα source and subsequently analyzed to acquire the chemical and structural information. The calculated reflectivity curve that best reproduces the experimental data is shown with its electron density depth profile plotted in Fig. 3b. When fitting the reflectivity curve of sample S1, a single layer model is used given the fact that the Fe seedlayer is very thin and its density is close to $Fe_{16}N_2$. At the bottom interface between the film and the substrate, an additional "transition layer" with increased electron density compared to MgO is detected, which *creates* the modulation of the oscillations in the reflectivity data and is consistently seen in modeling the data of sample S2. From the fit to the data with the bulk value of the electron density for MgO ($\sim 3.1\pm 0.1 \times 10^{-5}$Å$^{-2}$) we obtained a uniform layer of Fe$_{16}$N$_2$ ($5.9\pm 0.3 \times 10^{-5}$Å$^{-2}$), suggesting a uniform chemical composition normal to the surface. The introduction of two layer model for describing the XRR of sample S2 only marginally improves the quality of the fit.

Both samples were investigated by polarized neutron reflectometry (PNR) using the MAGICS Reflectometer at Spallation Neutron Source at Oak Ridge National Laboratoty (SNS ORNL)[20]. PNR allows the interface magnetism study[21], the *absolute* magnetization determination and magnetic depth profile in both simple thin films[22, 23, 24] and complicated superlattice structures[25, 26]. The reflectivities with the spin of the neutrons being either parallel (R+) or anti-parallel (R-) to the applied magnetic field were collected simultaneously. From these data the depth profiles of the scattering length density of both nuclear (NSLD) and magnetic (MSLD) are obtained. The PNR experiments were performed at room temperature in the saturation external field of H=1.0 T applied in-plane of the sample. The R+ and R- reflectivity data were fitted simultaneously using a genetic algorithm with an exact recursive matrix calculation embedded in the *Simulreflec 1.0* package[27]. In data modeling process, the structural NSLD profile was constrained to closely match x-ray results. To account for the possibility that the film possess homogeneous chemical composition but potentially different magnetization, the Fe-N layer was subdivided into three slabs where NSLD was fixed but thickness, roughness and MSLD were allowed to vary. Since all the parameters chosen for the top and bottom interfaces are preset in accordance to the x-ray results and are allowed for slight modulation during data analysis, the only free parameter is magnetic moment. The experimental reflectivity and calculated curves with best chi-squared fit for samples S1 and S2 are shown on Fig. 4a and b respectively. Their corresponding structural NSLD and magnetization depth profiles are plot in Fig. 4c and d. To compare XRR and PNR for the chemical SLD, the interface layer produces relatively large roughness in the PNR, which can be attributed to the lack of high q information or likely due to the non-magnetic nitrides (e. g. MgO:N) formed at the bottom interface accounting for the difference in scattering lengths for the neutron and x-ray probes on light elements such as N.

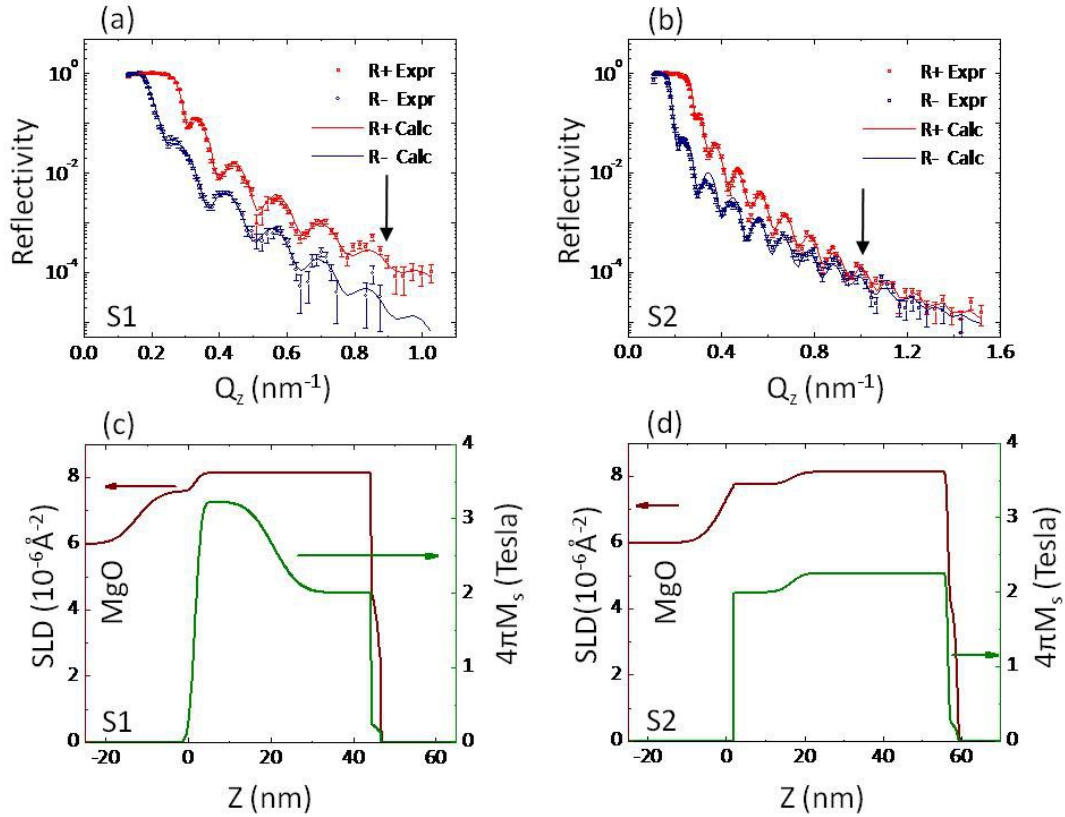

FIG. 4 **Polarized neutron reflectivity characterization** (a) and (b) Experimental polarized neutron reflectivites together with the fitted curves as functions of momentum transfer Q for sample S1 and S2 as labeled respectively. The arrows at high Q region (>0.8nm$^{-1}$) indicate the difference of the magnetic properties towards the bottom interface between these two samples (see text for detail). (c) and (d) structural (Brown) and magnetic (Green) depth profiles for samples S1 and S2 as labeled correspondingly.

It is clear that for sample S1, an anomalously large magnetization is present at the thickness range of about 20 nm towards the bottom interface, where MSLD is in the range of $7.2 \sim 7.5 \times 10^{-6} \text{Å}^{-2}$, corresponding to a magnetization of 3.1~3.2T, which is 40~50% larger than that of bulk Fe and 20~30% higher than that of $Fe_{65}Co_{35}$. As it approaches to the film surface, the MSLD drops rapidly and levels off at $4.66 \times 10^{-6} \text{Å}^{-2}$, corresponding to magnetization of ~2.01T, closely resembling that of nominal Fe. It is worth noting that a single layer model with both MSLD and NSLD to be uniform through the Fe-N layer or without introducing the transition layer failed to provide satisfying solutions (see auxiliary material Fig. SI2). For sample S2, following repeating analysis that co-refining PNR with XRR, the resulted MSLD is close to $5\times10^{-6}Å^{-2}$ for the Fe-N layer, corresponding to $M_s$ of 2.15T, which does not show the presence of giant $M_s$.

Further justification for the different magnetic structure upon Fe buffer thickness change comes from the spin asymmetry (SA) $(R^+-R^-)/(R^++R^-)$ plot shown in Fig. 5. In particular, the SA of the high $M_s$ sample (S1) and normal $M_s$ samples (S2) are plotted in the same scales. It is noticed that at high scattering vector ($q>0.8nm^{-1}$) region when both the MSLD

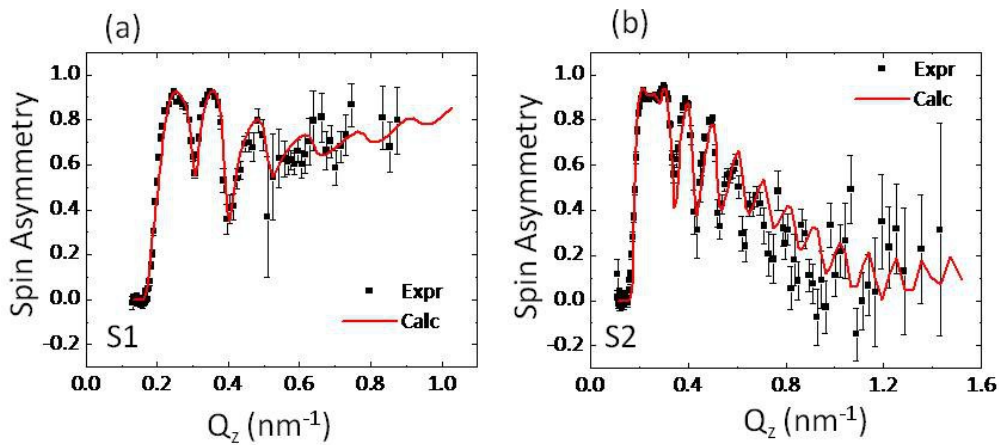

FIG. 5 **Spin Asymmetry analysis** (a) and (b) show the experiment (dots) and fitting (lines) data on sample S1 and S2, respectively. The difference of the high q ($q>0.8nm^{-1}$) behavior suggests the large magnetization developed in sample S1 but not in S2

and NSLD at the bottom interface dominate the behavior of R+ and R- reflectivities, in contrast to sample S2, the SA of the S1 shows a clear tendency to go to unity, which is reflected in the actual reflectivity curves as marked by arrows. This is only possible when condition |MSLD-NSLD|>>MSLD is satisfied. Since NSLD is similar for both samples, this observed feature directly proves the substantial enhancement of MSLD in S1 comparing to S2 at the bottom interface.

Given the large variation of the magnetization along the substrate-film normal and the disappearance of giant $M_s$ after introducing thicker Fe buffer, it is appealing to connect the straining effect to the formation of high $M_s$. It is known that by strained epitaxial growth, physical and magnetic properties can be altered significantly. Remarkable examples are that some materials exhibit ferroelectrocity[28] or anomalous ferromagnetism[29, 30] only in strained films. In common rigid metals, straining of crystal lattice by coherent growth is limited to ultrathin films with thicknesses of up to several atomic layers due to the requirement of substantial elastic energy. However, in the case of ferromagnetic martensite (Fe-N martensite in this case) that the energy scale is relatively flat over the entire Bain path[31], it is possible to stabilize the intermediate lattice geometry over a wide thickness range. The previous reported $M_s$ value on Fe-N epitaxial films sensitively depends on the choice of substrate, buffer layer and processing techniques (Ref. 12~16), though careful x-ray and electron diffraction results on these samples all show "similar" crystal structure of $Fe_{16}N_2$ phase as proposed by Jack et al[32], implying the subtle correlation between the magnetism and straining of the which is proved to be substantial as probed by GIXRD in the films discussed here.

On the other hand, the Slater Pauling (SP) curve (Ref. 4) describes the variation of average magnetic moment with the electron concentration in the framework of itinerant electron bands. To rationalize the observed $M_s$ which is beyond the SP curve, it is essential to require the electrons of interest to possess localized feature, where the itinerant ferromagnetism can no longer be suitably applied. In a view consistent with Hubbard's model, the presence of the crystal field ($\Delta$), the kinetic energy (k) and screened Coulomb interaction (U) determine the spin configuration of Fe atoms. It is known that an isolated Fe atom produces a magnetic moment of ~ 4.0 $\mu_B$ according to Hund's first rule[33]. When it comes to Fe metal, the large kinetic energy is compensated by the reduction of Coulomb U due to the strong screening produced by the electron itinerancy. Therefore, to enable the mobility and inter-atomic coupling, an intimidate spin state of ~ 2.0 $\mu_B$ per Fe is inevitable[34]. In $Fe_{16}N_2$, as previously alluded[35,36], the introduction of the N site provides unoccupied orbitals. As a result, the neighboring Fe sites transfer charges to the N site and facilitate electron conduction. In this case, both high spin (HS) and intermediate spin (IS) states are possible without significant modification of k, $\Delta$ and U in which the metallic property of the system is preserved. Therefore, in a special case when $|\Delta+k-U|\ll U$, a crystal field with small perturbation ($\Delta'$, $|\Delta-\Delta'|\ll\Delta$) and subsequent modification of kinetic energy (k') and Coulomb interaction (U') introduced by slight lattice distortion, can significantly influence the spin configuration of Fe sites, when conditions $\Delta+k-U>0$ and $\Delta'+k'-U'<0$ are spontaneously satisfied. In the case of $\Delta+k-U>0$, IS state is energetically favorable, which predicts a non-high $M_s$ scenario and is consistent with reports based on bulk samples in which the lattices are known to be fully relaxed[37]. However, in strained films presented here as well as those prepared by MBE (Ref. 8), Sputtering Beam and etc (Ref. 13), a modulation of the lattice constant is anticipated, in which the proposed ($\Delta'+k'-U'<0$) becomes possible and subsequently yields a giant $M_s$ as observed. Though local spin density approximation (LSDA) based calculation show insensitive dependence of the lattice constant[38] and favor an IS configuration, it is generally accepted that Coulomb U is underestimated in the LSDA, which may produce large discrepancy given the delicate requirement of the onset of the HS ground state.

The PNR work done at SNS, ORNL and the work done at UMN are sponsored by U.S. Department of Energy, Office of Basic Energy Sciences under No. DE-AC02-98CH10886. The authors would like to thank Prof. C. Leighton and Prof. B. I. Shklovskii for helpful discussion.

*N. Ji and V. Lauter contributed equally to this work

†email: jpwang@umn.edu